\documentclass[11pt]{article}
\usepackage{algorithmic}
\usepackage{algorithm}

\usepackage{url}
\usepackage{graphicx}
\usepackage{caption}
\usepackage{subfigure}
\usepackage{fullpage}

\begin{document}

\title{An Efficient Cell List Implementation\\for Monte Carlo Simulation on GPUs}

\author{
Loren Schwiebert\\
		{Dept.\ of Computer Science}\\
		{Wayne State University}\\
       	{Detroit, MI 48202}\\
       	\texttt{loren@wayne.edu}
\and
 Eyad Hailat\\
		{Dept.\ of Computer Science}\\
		{Wayne State University}\\
       	{Detroit, MI 48202}\\
       \texttt{eyad@wayne.edu}
\and
Kamel Rushaidat\\
		{Dept.\ of Computer Science}\\
		{Wayne State University}\\
       	{Detroit, MI 48202}\\
       \texttt{ed3457@wayne.edu}
\and  
\and
Jason Mick\\
       {Dept.\ of Chemical Engineering}\\
		{Wayne State University}\\
       	{Detroit, MI 48202}\\
       \texttt{dw2413@wayne.edu}
\and Jeffrey Potoff\\
       {Dept.\ of Chemical Engineering}\\
		{Wayne State University}\\
       	{Detroit, MI 48202}\\
       \texttt{jpotoff@wayne.edu}
       }
%

\maketitle
\begin{abstract}

Maximizing the performance potential of the modern day GPU architecture requires judicious utilization of available parallel resources. Although dramatic reductions can often be obtained through straightforward mappings, further performance improvements often require algorithmic redesigns to more closely exploit the target architecture. In this paper, we focus on efficient molecular simulations for the GPU and propose a novel cell list algorithm that better utilizes its parallel resources. Our goal is an efficient GPU implementation of large-scale Monte Carlo simulations for the grand canonical ensemble. This is a particularly challenging application because there is inherently less computation and parallelism than in similar applications with molecular dynamics. Consistent with the results of prior researchers, our simulation results show traditional cell list implementations for Monte Carlo simulations of molecular systems offer effectively no performance improvement for small systems~\cite{Brugge, Rice}, even when porting to the GPU. However for larger systems, the cell list implementation offers significant gains in performance. Furthermore, our novel cell list approach results in better performance for all problem sizes when compared with other GPU implementations with or without cell lists.

\end{abstract}



\section{Introduction}\label{sec:Introduction}

Graphics acceleration hardware has been commercially available to consumers since the early 1980s. The highly-parallel architecture has encouraged researchers to port many different applications to these devices. With the availability of the CUDA API~\cite{cuda+programming+guide6.0, Wittenbrink2011} writing code that runs on NVIDIA GPUs became a much easier task.  Although significant speedups by the GPU over serial implementations have been achieved for many applications, not all applications are easily ported to the GPU. In particular, problem domains with serial behavior such as Markov chain algorithms~\cite{Frenkel+Smit:02} often benefit little from the SIMD parallelism available on the GPU. In this class of problems, one simulation step depends on the results of the previous step. Furthermore, due to the nature of the problem, predicting the results of the previous step creates a bias and invalidates the sampling distribution of the algorithm.

The main goal of simulating molecular systems using Monte Carlo (MC) and molecular dynamics (MD) methods is to compute equilibrium properties of classical many-body systems or to estimate the average properties of systems with a very large number of accessible states. However, MC methods make it feasible to simulate open systems that MD cannot simulate, because the MD algorithms are not designed for systems that permit the addition or deletion of particles~\cite{Frenkel+Smit:02}. For example, MC can be applied to the grand canonical ensemble method, which is useful in adsorption studies, where the amount of material adsorbed is a function of the chemical potential and temperature of the reservoir with which the material is in contact. Moreover, in simulations of two phase systems, the properties of the system vary widely in the interfacial region, such as between the gas and adsorbent, which are a strong function of system size. Hence, it is necessary to simulate large systems to minimize the influence of the interface on the properties of the corresponding bulk phases~\cite{Frenkel+Smit:02, rice:1}.

Other simulation techniques, such as molecular dynamics, typically require significantly more computation for each step of the simulation, and so are better-suited for parallel implementation. Even so, some previous simulations have used the GPU to implement the MC method~\cite{doi:10.1021/ct200474j}. Their implementation depends on an embarrassingly parallel algorithm that runs several concurrent simulations with small systems of 128 particles. Instead, our work uses the energy decomposition method (farm algorithm), which enables us to support configurations with over a million particles.

In~\cite{Parallel-canonical-Monte-Carlo:Keeffe:2009}, a parallelization method for the canonical MC simulations via domain decomposition technique has been presented, where each domain can be assigned to a separate processor and multiple moves can be simulated in parallel. Interprocess communication is required only when moving particles near the edge of a domain, since this requires interactions between adjacent domains. To limit this communication, each domain is partitioned into three subdomains. The size of the middle subdomain is chosen as large as possible to minimize interprocess communication. Although well-suited for a multi-core CPU, this approach does not expose the fine-grained parallelism required for an efficient GPU implementation.

Each time a particle is displaced in, removed from, or inserted into the simulation region, energetic decomposition requires that pairwise energy be calculated between this particle and all other particles. A radial cutoff is typically chosen to reduce the execution time by limiting the calculation of inter-molecular forces to only those particles within the cutoff. The forces due to interactions with particles outside of the cutoff can be approximated using tail corrections.\footnote{This is a reasonable approximation because the atomic forces decrease at the rate of $O(d^6)$ or $O(d^{12})$ with the distance, $d$.} Since interactions within only a small radius are considered, it is possible to create either a cell list or a neighbor list to organize particles based on their relative locations and ignore particles that are beyond the cutoff. In this way, not only are the energy and pressure computations of more distant pairwise interactions avoided, but also the calculation of distances between these particles.

A number of researchers have proposed efficient techniques for maintaining these lists. One approach, the Verlet list~\cite{Verlet} or neighbor list, maintains a list of neighboring particles, particles within the radial cutoff, for each particle. This minimizes the number of interactions that must be computed, but requires frequent updating. To reduce the frequency of updating, the Verlet list can be expanded to include particles slightly beyond the cutoff, so as particles move they do not enter or exit another particle's neighbor list until $n$ simulation steps. Although this is a reasonable option for MD simulations, where all the particles move with each step, this is not practical for MC simulations. In MC simulations, typically only a single particle is moving in each step and the displacement of this particle can be arbitrarily large. In addition, inserting or deleting a particle from the simulation, which is not possible in MD simulations, would require rebuilding the Verlet list after this move.

Another approach is to use a cell list, where the simulation box is partitioned into cells (squares in 2D, cubes in 3D) such that each cell has dimensions not much larger than the cutoff and perhaps smaller. The only particles that need to be considered are those in the same cell and the adjoining cells. For larger systems, as the simulation box grows relative to the cutoff, an increasing portion of the particles can be ignored. Because many of the particles in adjacent cells will be outside the cutoff, more potential pairwise interactions are computed than with a Verlet list. However, a cell list can be maintained with less overhead than a Verlet list. To reduce the number of extraneous particles processed in the cell list implementation, one can sort the particles in each cell~\cite{Grest1989, Yao}. Or the cells can be made smaller, with more cells processed for each move~\cite{Allen, Rice}; this can be taken to the extreme of a cell size large enough for only a single particle. However, this requires many more cells to be examined, many of which will be empty, which also introduces some overhead. Since the cell list approach is more promising for MC simulations on the GPU, for the rest of this paper we consider only cell lists.

A third option is to use both a Verlet list and a cell list~\cite{Grest1989}. For instance, Proctor \emph{et al.}~\cite{Proctor2013} show cell lists on the GPU allow a fast approximation of whether or not two particles are within the cutoff, which performs better than immediately traversing the neighbor list. They do not create or maintain a cell list, but calculate the cell of each particle based on its coordinates, with cell dimensions larger than the cutoff, and use this calculation to determine whether or not two particles are in neighboring cells.

There are many examples of using a cell list implementation for the MD simulations~\cite{Anderson+LorenzETAL-Genepurpmoledyna:08a, Daly20122054, Frenkel+Smit:02, stone2007accelerating, van2008harvesting}. On early GPUs, an efficient implementation of cell list on the GPU was not viable due to the lack of atomic operations on the GPU~\cite{doi:10.1021/ct200474j}. Instead, implementations such as~\cite{Anderson+LorenzETAL-Genepurpmoledyna:08a, stone2007accelerating, van2008harvesting} use the CPU to construct the cell list and then copy it to the GPU. These cell lists are then used to construct a neighbor list. Note that in molecular dynamics simulations, all molecules are moved in each step, requiring the cell list to be updated after nearly every simulation step. The frequency of updates depends on how far a molecule moves in each step, how much extra distance beyond the cutoff is used in defining the neighbors, and how much inaccuracy can be tolerated in the computations. A state-of-the-art implementation is described in~\cite{hoomd-blue2012}.

In MD simulations, since all the particles of the system are moving at the same time, the overhead of maintaining the cell list can be eclipsed by the computation cost of each step of the simulation. On the other hand, in a grand canonical MC simulation, only one particle is moved in each step. So, there is less computation. Therefore, prior work on cell lists for MC simulation showed that there is no performance gain for these applications with relatively small systems~\cite{Brugge, Rice}. Since MC simulations typically have run on systems with at most a few thousand particles, cell lists have been considered impractical for MC simulations.

In this paper, we reexamine this assumption for MC simulations on the GPU. There are two aspects to this reconsideration. First, with the GPU, it is feasible to simulate much larger systems in a reasonable amount of time. So, it is appropriate to ask whether cell lists might offer a performance advantage for much larger systems. Second, the GPU is a massively multithreaded architecture, so we evaluate a novel cell list implementation designed specifically for a manycore architecture such as the GPU.

To develop our code, we started with a state-of-the-art sequential code developed by Chemical Engineers in our research group to ensure that scientifically valid results are produced. The GPU versions were then compared with the output of this sequential code to ensure that not only fast code, but also correct code was produced. Our purpose is not, however, to compare the sequential code with the GPU code. Instead, we look at various versions of the GPU code, focusing on different cell list implementations to investigate the best cell list implementation for the GPU. Our results clearly demonstrate that a cell list implementation tailored to a manycore architecture offers significant speedup over both the original GPU code and the traditional cell list for problem sizes ranging from 512 particles up to over one million particles.

The rest of this paper is organized as follows. Section~\ref{subsec:GCMC} introduces the MC simulation for the grand canonical algorithm. Then, in section~\ref{sec:method} we introduce techniques, optimizations, and decisions we have made to write the parallel algorithms with and without the cell implementation. Both a traditional cell list implementation and our novel cell list for the GPU are presented. Performance results and observations are discussed in section~\ref{sec:discussion}. Section~\ref{sec:conclusion} presents our conclusions and future work.

\section{Grand Canonical Ensemble}\label{subsec:GCMC}

The grand canonical ensemble extends the canonical ensemble by fixing the values of the temperature ($T$), volume ($V$), and the chemical potential ($\mu$)~\cite{Frenkel+Smit:02}. Particles can interact with each other only when they exist inside the simulation box and are within a cutoff radius, $r_{cut}$, of each other. A reservoir is connected to this simulation box, allowing the particles and energy to be exchanged freely between them. Through this exchange of particles, the system and the reservoir will reach the equilibrium state, which can be determined by using the values of the temperature and the chemical potential.

This method can be applied to problems such as:
\begin{enumerate}

\item Simulating adsorption isotherms.  While it is essential to have detailed knowledge of the behavior of the adsorbed molecules, this type of information is very difficult to obtain experimentally; simulation is the alternative.

\item In numerical simulations to accurately predict properties of materials and their guest-adsorption characteristics.

\item Determining the equation of state of the Lennard-Jones fluid. One fixes the temperature and chemical potential and calculate the resulting density and pressure.
\end{enumerate}

We study systems of particles interacting via the Lennard-Jones potential by calculating the configurational energy of pairwise interaction, given by:
\begin{equation}\label{equ:ljU}
U(r)= 4\epsilon \left[ \left(\frac{\sigma}{r} \right)^{12} - \left(\frac{\sigma}{r} \right)^{6} \right],
\end{equation}
where $r$ is the distance between two interacting particles, $\epsilon$ is the depth of the potential well, and $\sigma$ is the particle diameter. Note that calculations of the Lennard-Jones potential are significantly more complex than the Ising or hard sphere models, since the system must calculate the interactions between all particles within a certain cutoff radius. On the other hand, the hard sphere system just requires checking for overlap.  In the Ising model, we are only calculating interactions between nearest neighbors and those nearest neighbors are always known, since the spins do not change locations during the simulation.

Particles in this ensemble are moving inside the simulation box or between the box and the reservoir. The acceptance of a move is determined by computing the Boltzmann factor:
\begin{equation}\label{equ:Boltzmannfactor}
B_F = e^{-\beta \Delta E},
\end{equation}
where $\beta$ is given by $\displaystyle{\left(1/{k_{B}T}\right)}$, $k_{B}$ is the \textit{Boltzmann constant}, and $T$ is the temperature of the system. $\Delta E = [U(s^{\prime N}) - U(s^N)]$ is the difference between the new system energy and the old one, where $U$ is the total energy of the system for a given configuration, $N$ is the number of particles in the box, and $s$ and $s^{\prime}$ represent the old and new positions of the particle, respectively.

The acceptance or rejection for the types of particle moves are given by the Metropolis acceptance criterion~\cite{metropolis:1087, TheMetropolisAlgorithm-Beichl:2000}: %
\begin{description}
\item[Particle Displacement] A random particle is attempting to move randomly within the simulation box. The move is accepted with a probability:
\begin{equation}\label{equ:atomdisp}
acc(s \rightarrow s^{\prime}) = \min\left[1, B_F \right]
\end{equation}

\item[Insertion] A particle is inserted from the infinite reservoir into a random location in the simulation box. The acceptance probability of this move is giving by:
\begin{equation}\label{equ:Insertion}
acc(C \rightarrow C+1) = \min\left[\ 1,\frac{V B_F}{\Lambda^3 (N+1)} \right],
\end{equation}
where $\Delta E$ equals $[\mu - U (N+1)+U(N)]$ and $\Lambda$ is the \textit{thermal de Broglie wavelength}.
\item[Deletion]
The transfer of a random particle out of the simulation box into the reservoir is accepted with a probability:
\begin{equation}\label{equ:Deletion}
acc(C \rightarrow C-1) = \min\left[1, \frac{\Lambda^3N B_F}{V}\right],
\end{equation}
where $\Delta E$ in this case equals $[\mu + U(N-1)-U(N)]$.
\end{description}

Algorithm~\ref{alg:serialgc} shows how the serial code works. In general, this algorithm executes \emph{DisplacePercent}\/ of the simulation steps as particle displacement moves, and the rest are divided equally between the insertion and deletion of particles.

In each of the three moves, the computational cost is dominated by the overhead to calculate the pairwise energy with all interacting particles. Since this is a Markov chain algorithm, each step needs the current system status to calculate the probability of acceptance for the next one.

\begin{algorithm}[!t]
\caption{Serial Grand Canonical Ensemble Monte Carlo Algorithm}
\label{alg:serialgc}
\algsetup{indent=2em}
{\footnotesize
\begin{algorithmic}[1]
\STATE \textbf{Input:} One box of size (N) particles, (V) volume
\STATE \textbf{Input:} Infinite reservoir
\STATE //Initialize N particles positions inside the box
\STATE //Calculate total system energy
\STATE //Main simulation loop
        \FOR { each step }
      	\STATE //Randomly select a move type
		\STATE  \textit{R} $\leftarrow$  {\bf rand}() 
	\label{line:randommove}
	\IF {(\textit{R} $ < $ DisplacePercent)}
		\STATE //Attempt particle displacement
		\ELSE
			\STATE //Attempt particle transfer
			\STATE //Insertion/Deletion
			\STATE //Chose a random source of particle
		 	\STATE 	\textit{Source} $\leftarrow$ {\bf rand}()
		\IF { (\textit{Source} $<$ 0.5)}
			\STATE //Source box is the box (Deletion)
		\ELSE
			\STATE //Source box is the reservoir (Insertion)
        \ENDIF
		\ENDIF
		\STATE //Solve if the system in equilibrium (Balance)
		\STATE //Periodically write system status to disk
        \ENDFOR

\end{algorithmic}
}
\end{algorithm}

\section{GPU Implementations}\label{sec:method}

Due to the highly multithreaded architecture of graphics devices, fine-grained parallelism is needed to keep the processors in the device busy. For example, on NVIDIA GPUs, threads are scheduled in warps of 32 threads. For good performance, all 32 threads in a warp must execute the same instruction and avoid branches in the code that lead to warp divergence. Another requirement to achieve good performance is to hide the memory latency. Even though the GPU has high memory bandwidth, the relatively high latency of global memory accesses has to be hidden through sufficient parallelism to get efficient performance.

\subsection{Implementation without Cell List}

Although the simulation steps have been implemented on the GPU, the CPU makes the decision on which move to execute next and calls the corresponding kernel. Moreover, the simulation periodically writes system status to disk, an operation not supported by current GPUs.

In this section, we describe in detail the simulation moves implemented on the GPU without the use of cell lists. Our implementation of the traditional cell list and our new microcell list are presented in sections~\ref{subsec:cellListImp} and~\ref{subsec:microcellListImp}, respectively.

\subsubsection{Calculating Total System Energy}\label{subsubsec:CTE}

Although this is the most time consuming kernel call, since all pairwise interactions are being calculated, this function is called only once to calculate the initial system energy. The total number of unique pairwise energy calculations is potentially $N(N-1)/2$. Rather than create one thread for each pair, we create $4N$ threads and use these threads to iterate over all unique pairs of particles using a method akin to the RB technique~\cite{Navarro}. Since this is computed only once, minimal effort was expended in optimizing this function.

After each thread finishes calculating the pairwise energy, it writes the results to shared memory. A reduction operation is then executed on all threads in a block. Afterward, the summation of the thread values in the block is transferred to global memory to be visible to other blocks. The last block to finish copies the partial sums from all other blocks into its shared memory. Another iteration of reduction for these sums is executed in shared memory and the result is copied to global memory to be used by other kernel calls.

%

\subsubsection{Particle Displacement within the Box}

In this move, a randomly selected particle attempts to move to a random position within the simulation box. The amount of energy that this particle is contributing to the system in the new location should be calculated, by first deducting the particle's energy contribution from its original location, then calculating the additional energy for the new location. Since only one particle is moved, this approach allows us to update the system energy using an $O(n)$ operation. The reduction technique described in section~\ref{subsubsec:CTE} is used here. The difference in energy, $\Delta E$, is used for calculating the Boltzmann factor in equation~\ref{equ:atomdisp} to decide whether or not to accept the new position. Upon acceptance of the particle displacement, the system status is updated, which includes the new energy, new virial pressure, particle's new  position, and other configuration variables.

\begin{algorithm}[!t]
\caption{Parallel particle displacement}
\label{alg:trymove}
\algsetup{indent=2em}
{\footnotesize
\begin{algorithmic}[1]
\STATE \textbf{Input:}  One box of size (N) particles, (V) volume
       	\STATE //Randomly select a particle to displace
		\STATE  \textit{P} $\leftarrow$  {\bf rand}() 
		\STATE  $\Delta E \leftarrow$ CalculateParticlesContributionTM()
		\IF{thread 0 in last block}
		\STATE Use $\Delta E$ to calculate the acceptance rule
		\STATE //Select a random number \textit{A} in [0,1)
		\STATE  \textit{A} $\leftarrow$  {\bf rand}()
		\IF{A $<$ ProbOfAcceptance}
			\STATE  //Move accepted, apply changes
			\STATE //Update cell contents
		\ELSE
		  \STATE  //Move rejected
		\ENDIF
		\ENDIF
\end{algorithmic}
}
\end{algorithm}

In algorithm~\ref{alg:trymove}, the kernel function {\sf TryMove()} calls a GPU device function, {\sf CalculateParticlesContributionTM()}, that returns $\Delta E$. Thread zero of the last block to finish decides whether or not to accept the move by computing the probability of acceptance using $\Delta E$ and comparing this with a random value. If the move is accepted, then the new particle position, current energy, etc.\ are updated. This information is maintained on the GPU and only copied to the CPU for periodic configuration dumps. Algorithm~\ref{alg:CalculateParticlesContributionTM} shows the parallel algorithm for the function {\sf CalculateParticlesContributionTM()}.

\begin{algorithm}[!t]
\caption{CalculateParticlesContributionTM \\(no cell list)}
\label{alg:CalculateParticlesContributionTM}
\algsetup{indent=2em}
{\footnotesize
\begin{algorithmic}[1]
\STATE \textbf{Input:}  One thread per particle
\STATE //Initialize shared memory
\STATE //Assign a particle for the current thread
\FOR{the ParticleID corresponding to this thread}
	\STATE  //For the particle in the old location
	\STATE  //Determine the true distance between particles,
	\STATE	// applying periodic boundary conditions.
	\STATE //Calculate RadialDistance between particles.
	\IF{RadialDistance within cutoff}
		\STATE //store interaction results in shared memory
	\ENDIF
		\STATE  //For the particle in the New location
		\STATE  //Determine the true distance between particles,
		\STATE	// applying periodic boundary conditions.
		\STATE //Calculate RadialDistance between particles.
		\IF{RadialDistance within cutoff}
			\STATE //store interaction results in shared memory
		\ENDIF
\ENDFOR
\STATE syncthreads()
\STATE //Apply reduction in shared memory
\STATE //Move results from each block to global memory
\STATE //Last block moves data from global to shared memory
\STATE //Apply reduction in shared memory
\STATE //final result is  $\Delta E$
\STATE //return $\Delta E$ to caller via global memory
\end{algorithmic}
}
\end{algorithm}

\subsubsection{Insertion and Deletion of Particles}

To insert a particle from the infinite reservoir, a random position is selected in the simulation box. The device function {\sf CalculateParticlesContributionTPT()} calculates the energy contribution for the new particle in its new location. The change in system energy due to this insertion is calculated by assigning one thread for each particle. The result of each pairwise energy interaction is stored in shared memory. The reduction process described in section~\ref{subsubsec:CTE} is then executed as illustrated in algorithm~\ref{alg:CalculateParticlesContributionTPT}.  Algorithm~\ref{alg:insertion} shows the next steps of calculating the probability of acceptance, generating a random number, and comparing the results. If the move is accepted, the current system parameters will be updated to reflect the addition of this particle; otherwise, the system configuration remains unchanged. Since each kernel is called with one thread per particle, the number of particles in the box needs to be copied from the GPU to the CPU after each insertion or deletion move.

The deletion move is similar to the insertion move, except that the system is removing a particle and has to change the probability of acceptance as in equation~\ref{equ:Deletion}. When the deletion step is accepted, the particle moves to the reservoir and the system configuration is updated accordingly.

\begin{algorithm}[!t]
\caption{Calculate Particles Contribution (no cell list)}
\label{alg:CalculateParticlesContributionTPT}
\algsetup{indent=2em}
{\footnotesize
\begin{algorithmic}[1]
\STATE \textbf{Input:} One thread per particle
\STATE //Initialize shared memory
\STATE //Assign a particle for the current thread
	\STATE  //For all other particles in the box
	\STATE  //Determine the true distance with ParticleID,
	\STATE	// applying periodic boundary conditions
	\STATE //Calculate RadialDistance between particles
	\IF{RadialDistance within cutoff}
		\STATE //Store interaction results in shared memory
	\ENDIF

\STATE syncthreads()
\STATE //Apply reduction algorithm
\STATE //return $\Delta E$ to caller via global memory
\end{algorithmic}
}
\end{algorithm}

\begin{algorithm}[!t]
\caption{Parallel Insertion/Deletion}
\label{alg:insertion}
\algsetup{indent=2em}
{\footnotesize
\begin{algorithmic}[1]
\STATE \textbf{Input:} One box of size (N) particles, (V) volume
\STATE \textbf{Input:} An infinite reservoir
       	\STATE //Randomly select a position to insert into the box
       	\STATE //Find designated cell
	\STATE //Calculate the new particle's energy contribution
		\STATE  $\Delta E \leftarrow$ CalculateParticlesContributionTPT()
		\IF{thread 0 in last block}
		\STATE //Use $\Delta E$ to calculate the acceptance rule
		\STATE //Select a random number \emph{A} in [0,1)
		\STATE  \emph{A} $\leftarrow$  {\bf rand}()
		\IF{A $<$ ProbOfAcceptance}
			\STATE  //Move accepted, apply changes
		\ELSE
		  \STATE  //Move rejected
		\ENDIF
		\ENDIF
\end{algorithmic}
}
\end{algorithm}

\subsection{Traditional Cell List Implementation}\label{subsec:cellListImp}

\begin{figure}
\centering
\includegraphics[width=0.7\linewidth]{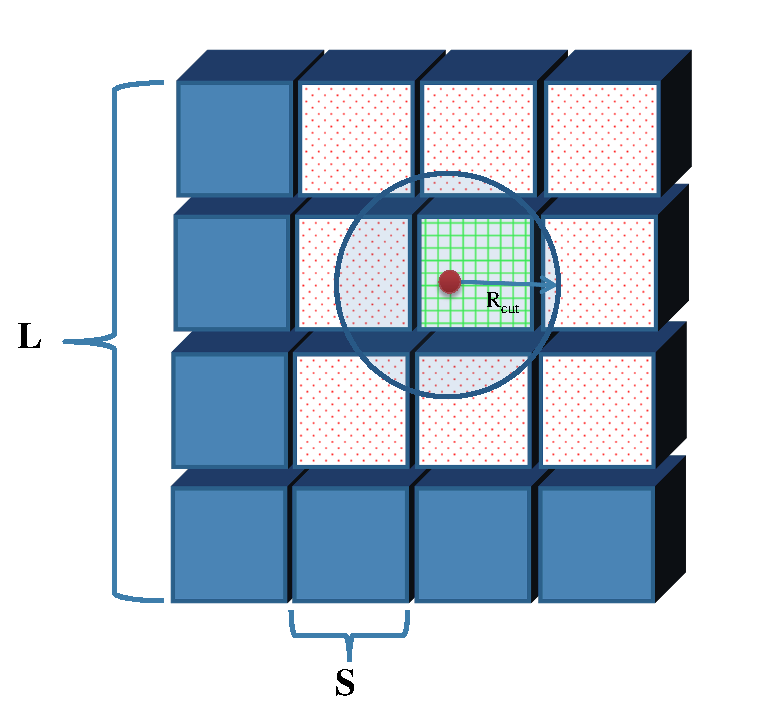}
\caption{The volume $V=L^3$ is decomposed into $T \ge 3$ cells per dimension, with cell dimension of size $S \ge r_{cut}$.}
\label{fig:cellsize}
\end{figure}

\begin{figure}
\centering
\includegraphics[width=0.7\linewidth]{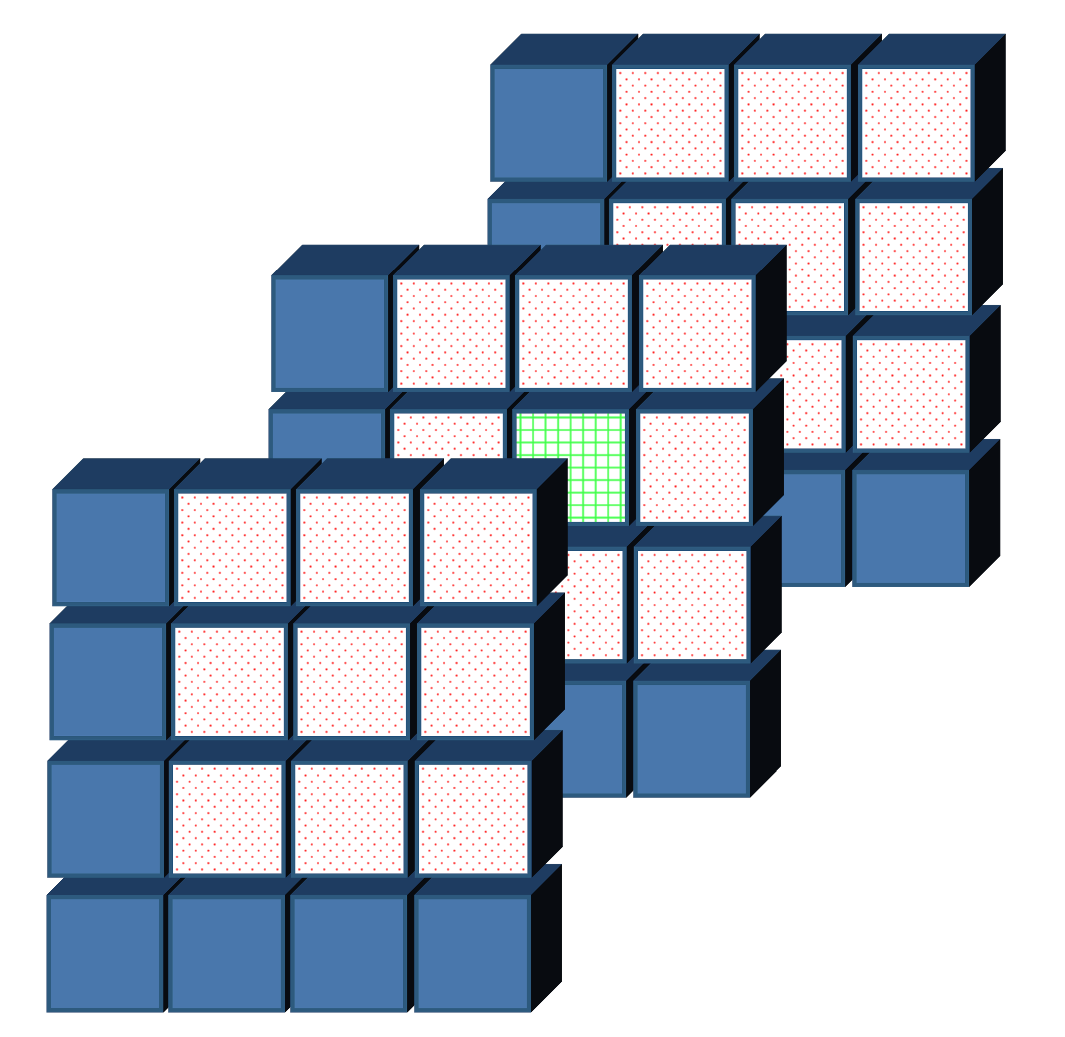}
\caption{26 cells adjacent to a particle in all axes.}
\label{fig:adjacent_cells}
\end{figure}

As depicted in figure~\ref{fig:cellsize}, if we use a short radial cutoff, denoted $r_{cut}$, which is typical,\footnote{A cutoff of $2.5\sigma$ to $3.0\sigma$ is not uncommon.} the simulation box can be decomposed into smaller domains, called cells, with the cell length along each dimension $S$ greater than or equal to $r_{cut}$. For a given particle, all interacting particles are located in the same or adjoining cells along all axes. Figure~\ref{fig:adjacent_cells} shows a 3D model of a simulation box where the dotted cells are the neighboring cells to the cell with grid lines. Each move has a constant upper bound on the number of particles that need to be considered for pairwise interactions, which is dependent on $r_{cut}$ but independent of the system size. This property holds because of the physical reality underlying our simulation, which constrains the minimum distance between particles.

On the other hand, the cell list algorithm suffers from the associated overhead of constructing, storing and accessing, and maintaining the cell structure. For small systems, this overhead may exceed the advantages of the cell list because only a few particles are considered even without cells. In addition, systems with a small box or a large cutoff radius cannot be decomposed into more than three cells per dimension. So, the entire simulation box is included within the adjoining cells, but the overhead of the cell list is also present. Next, we discuss the algorithm for implementing the traditional cell list and various factors of the design.

\subsubsection{Building the Initial Cell List}

Many techniques have been proposed for implementing cell lists~\cite{Frenkel+Smit:02}. On the GPU, we must use a method that exposes a great deal of parallelism. Some approaches use a linked list to store the indices of the particles in each cell, while others assign a fixed sized array of placeholders to every cell. A significant disadvantage with a linked list is that parallel access to the particle indices is not possible. The disadvantage of the latter scheme is the extra memory that may be wasted. Although our implementation uses the latter scheme, experiments show that even for large systems, the maximum number of particles in a cell can be relatively small when a suitable cell size is chosen. This is discussed in more detail in section~\ref{subsubsec:cellsize}.

In this implementation, we applied cell lists to run entirely on the GPU, constructed once at the beginning of the simulation and requiring minimum maintenance. First, a data structure (\emph{ParticlesInCells}) is constructed to hold the 26 adjacent cell IDs for each cell, of size $27\,T$, where $T$ is the number of cells in the box. Then, \emph{ParticlesInCells} is bound to texture memory to take advantage of caching. Later for the entire simulation run, values will be read through texture fetches. Using texture memory is efficient for this purpose, especially since one thread block will be reading the same cell indices for the entire kernel call, allowing for a high rate of cache hits. Algorithms~\ref{alg:CalculateParticlesContributionTMCell} and~\ref{alg:CalculateParticlesContributionTPTCell} show how the cell list implementation is used for finding $\Delta E$ for both particle displacement and particle insertion/deletion, respectively.

The process of placing particles in cells is called \emph{binning} the particles. The binning process involves looping through the $N$ particles and placing each into one of the $T$ cells. The particle binning is done when we first construct the simulation box and generate the initial locations of all particles. When a particle is added to a cell, an atomic operation is performed on the relevant cell counter in order to avoid a race condition, so that each particle added to a cell is placed in a unique array location. Since atomic operations have become faster with the Fermi and Kepler architectures, and since this is done just when the system configuration is initialized, the overhead is minimal.

\begin{algorithm}[!t]
\caption{CalculateParticlesContributionTM function (cell list $9\times 3$)}
\label{alg:CalculateParticlesContributionTMCell}
\algsetup{indent=2em}
{\footnotesize
\begin{algorithmic}[1]
\STATE \textbf{Input:} Three cells per block
\STATE //Initialize shared memory (x3)
\STATE //For the old position of the particle
\STATE //Find SourceCurrentCell
\STATE //Fetch three neighboring cells from texture
\IF{Non-empty cell}
\STATE //For each ParticleID in a fetched cell
	\STATE //Assign one thread per particle
	\STATE //Determine the true distance between particles,
	\STATE // applying periodic boundary conditions.
	\STATE //Calculate RadialDistance between particles.
	\IF{RadialDistance within cutoff}
		\STATE //store interaction results in shared memory
	\ENDIF
\ENDIF
\STATE //For the new position of the particle
\STATE //Find DestCurrentCell
\STATE //Fetch three neighboring cells from texture
\IF{Non-empty cell}
\STATE //For each ParticleID in a fetched cell
		\STATE //Determine the true distance between particles,
		\STATE // applying periodic boundary conditions.
		\STATE //Calculate RadialDistance between particles.
		\IF{RadialDistance within cutoff}
			\STATE //Store interaction results in shared memory
		\ENDIF
\ENDIF
		
\STATE syncthreads()
\STATE //Apply reduction algorithm
\STATE //return $\Delta E$ to caller via global memory
\end{algorithmic}
}
\end{algorithm}

\begin{algorithm}[!t]
\caption{CalculateParticlesContributionTPT function (cell list $9\times 3$)}
\label{alg:CalculateParticlesContributionTPTCell}
\algsetup{indent=2em}
{\footnotesize
\begin{algorithmic}[1]
\STATE \textbf{Input:} Three cells per block
\STATE //Initialize shared memory (x3)
\STATE //For the selected particle (deletion) or
\STATE // selected position (insertion)
\STATE //Find CurrentCell
\STATE //Fetch three neighboring cells from texture
\IF{Non-empty cell}
\STATE //For each ParticleID in a fetched cell
	\STATE //Assign one thread per particle
	\STATE //Determine the true distance between particles,
	\STATE // applying periodic boundary conditions.
	\STATE //Calculate RadialDistance between particles.
	\IF{RadialDistance within cutoff}
		\STATE //Store interaction results in shared memory
	\ENDIF
\ENDIF
		
\STATE syncthreads()
\STATE //Apply reduction algorithm
\STATE //return $\Delta E$ to caller via global memory
\end{algorithmic}
}
\end{algorithm}

\subsubsection{Cell Size}\label{subsubsec:cellsize}

Using the notation in figure~\ref{fig:cellsize}, we select $S$ to maximize the integer $T = L/S$ with the constraint that $S \ge r_{cut}$. For instance, if $r_{cut} = 2.5$ and $L = 23.9$, then $S = 2.656$ with $T = 9$ cells per dimension. If $L < 3\,r_{cut}$, we set $S = L/3$. The use of periodic boundary conditions means that the entire simulation box will always be contained in these $27$ cells, so cells smaller than $r_{cut}$ are acceptable.

A fixed size array of placeholders for every cell has been used to support cell lists in our code. This parameter depends on the density of the simulation box, the radial cutoff, and the minimum distance between particles. A larger array size means extra wasted memory locations. On the other hand, a small array size might prevent valid configurations from being realized. Our implementation considers an array size large enough to encompass all particles within range to avoid overflow. The code also outputs an error message if array overflow would occur and the simulation must then be rerun with a larger array size. To avoid this problem, we selected sufficiently large array sizes through experimentation. The number of particles in a cell for a simulation with a small cutoff can be limited to 48 particles per cell. For cutoffs greater than $4.0\sigma$, we require a limit of 96 particles per cell.

Note that move attempts where a particle stays in the same cell can be processed more efficiently. Whenever the particle moves within the same cell, the neighboring cells are the same for both the old and new locations for this particle. Hence, calculating the pairwise energy for both the old and the new location will consider the same set of particles. The coordinates of these particles are cached when calculating the energy at the old location and these cached coordinates can be reused when recalculating the energy at new location. Furthermore, if the move is accepted, since the list of particles in the cell is unchanged, no updates need to be made to any cells.

\begin{figure*}
\centering
\subfigure[One cell per block]{\label{fig:27x1}\includegraphics[width=3.4in]{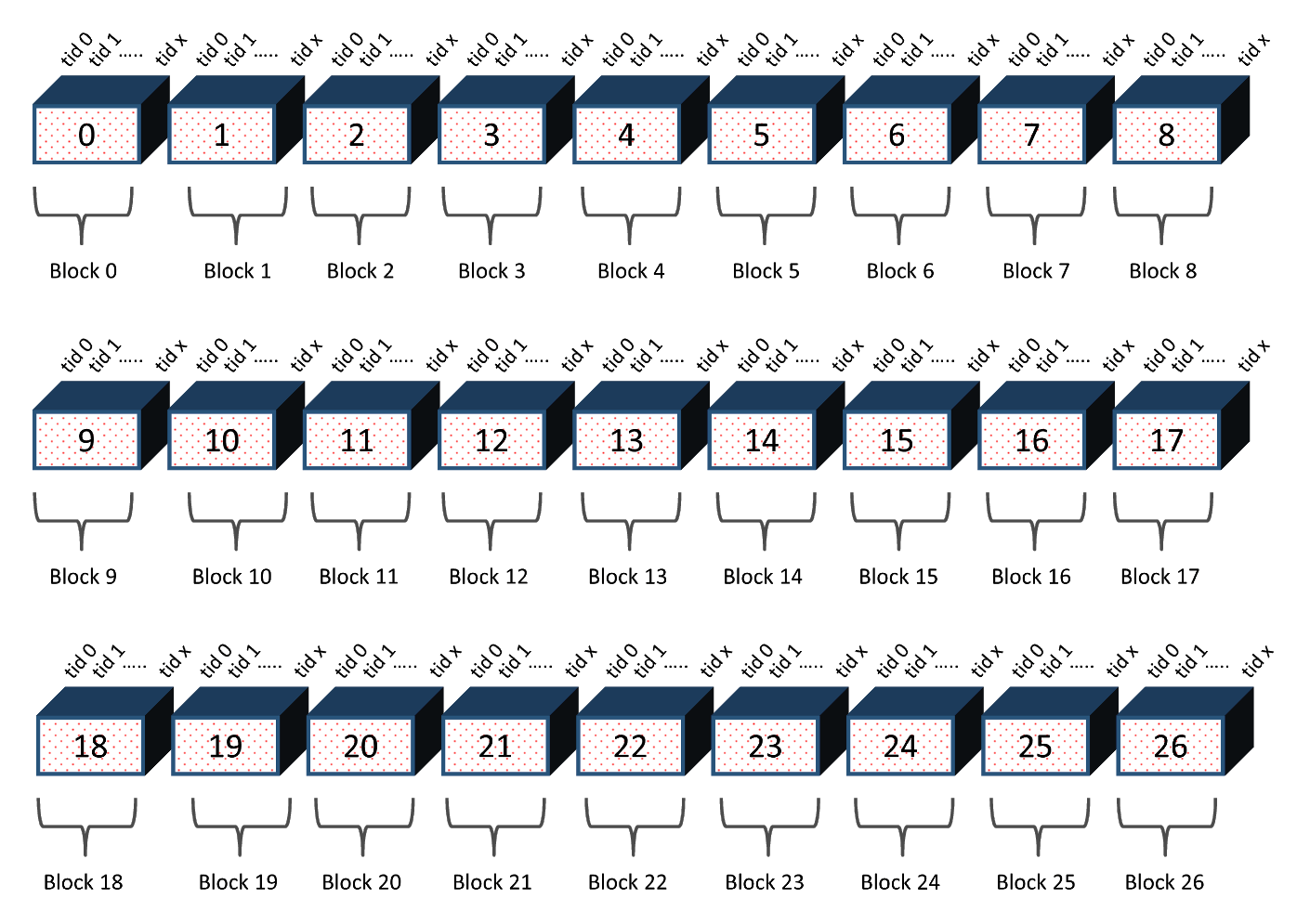}}
\subfigure[Nine cells per block]{\label{fig:3x9}\includegraphics[width=3.4in]{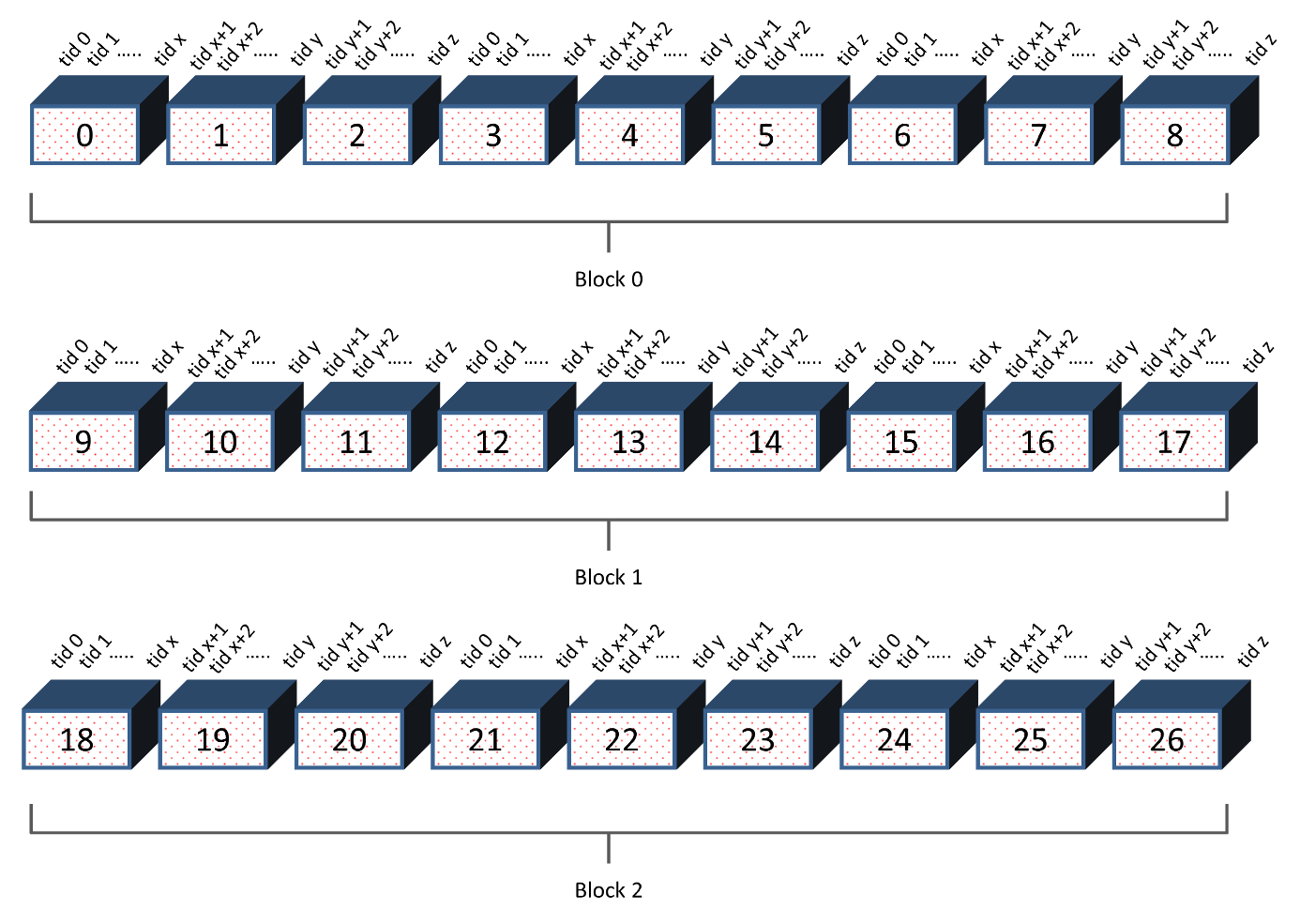}}
\subfigure[Three cells per block]{\label{fig:9x3}\includegraphics[width=3.4in]{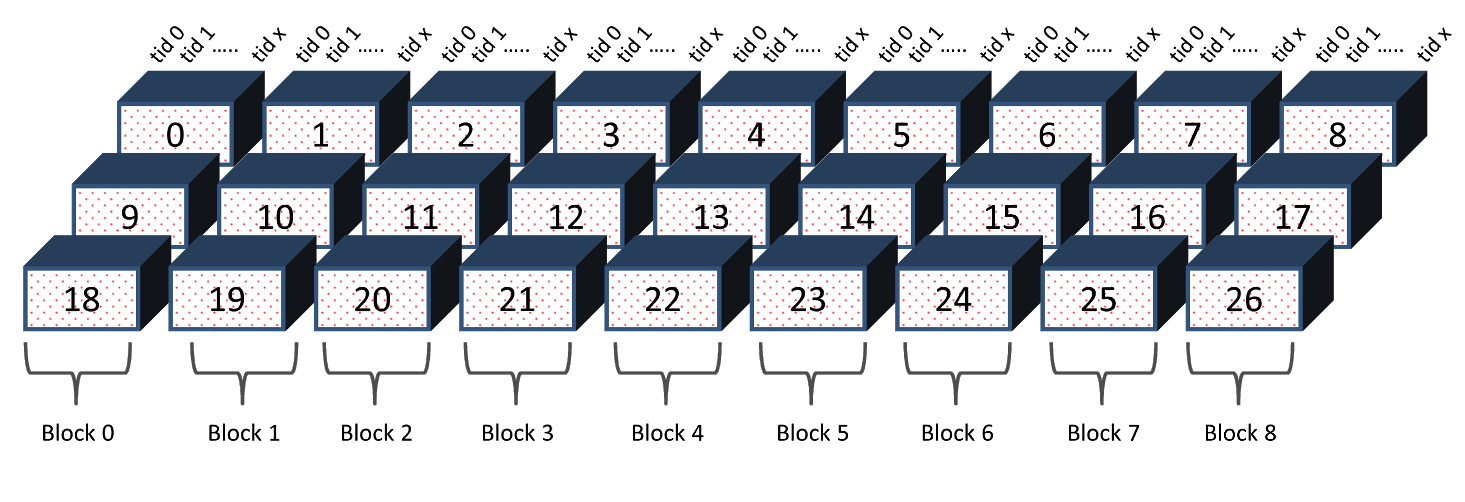}}
\subfigure[Twenty-seven cells per block]{\label{fig:1x27}\includegraphics[width=3.4in]{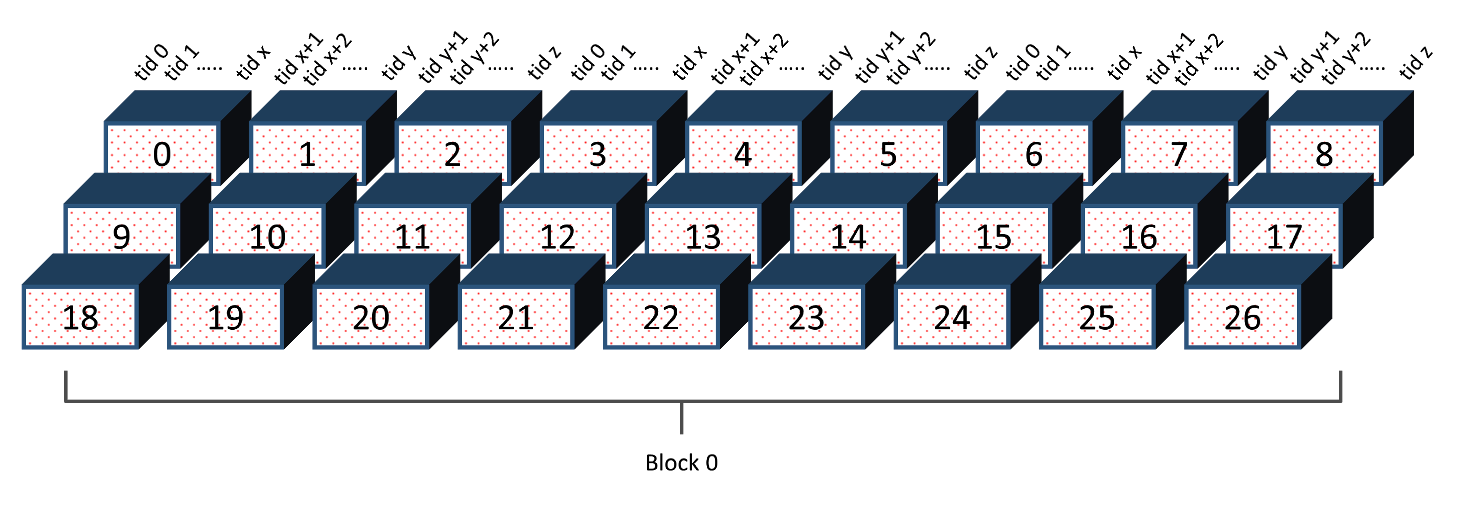}}
\caption{Different methods of assigning cells to thread blocks.}
\label{fig:cellsperblocks}
\end{figure*}

\subsubsection{Assigning Cells to Blocks} \label{subsec:assigningCellstoBlocks}

To the best of our knowledge, previous implementations of cell lists on the GPU have used a mapping where each cell is assigned to a different thread block. In MD simulations, where all particles are moved at the same time, it is more efficient to assign one thread per particle and one block per cell to take advantage of caching all 26 neighboring cells. However, this is not necessarily true for MC simulations where only one particle moves in each simulation step, and more blocks need more synchronization through atomic operations on global memory. Our implementation considers multiple options, with differing numbers of cells per block and reports the performance differences. This is depicted in figure~\ref{fig:cellsperblocks}, where each thread block may compute 1, 3, 9, or 27 cells.

\begin{description}
\item[One Cell per Block]
This is the most straightforward implementation expressed in figure~\ref{fig:27x1}. In this scheme, a kernel launches with 27 blocks, and the block size is set as the number of particles per cell.  However, more synchronization between blocks and more total shared memory is needed. Yet, for systems with a uniform distribution of particles per cell, this works best.

\item[Three and Nine Cells per Block]
Due to the nature of the MC simulation, only one particle is examined each simulation step. This leads to relatively low device utilization. In figures~\ref{fig:3x9} and~\ref{fig:9x3},  two different ways of GPU resource management are considered, one with nine cells per block and the other with three cells per block, respectively. In both cases, a separate thread is assigned to each particle. These two options use more total local memory per kernel call, with less need for block synchronization.

\item[27 Cells in One Block]
Figure~\ref{fig:1x27} shows the option of assigning all 27 cells to one block. Due to the many threads in this block with this alternative, more shared memory and other GPU resources are required. This also means that only one SM is used to compute the entire simulation and the other SMs are idle. This option is useful if one wishes to run several large simulations concurrently.

\end{description}

\subsubsection{Assigning Threads to Particles}

Previous work has studied the performance of one cell per block and one thread per particle~\cite{stone2007accelerating, van2008harvesting}. However, for very large systems that were too large to be simulated prior to this work, it is worthwhile to study the performance of assigning multiple particle per thread. An algorithm has been developed to assign multiple particles from more than one cell to the same thread. For example, for a kernel of nine blocks, if each block has been called with one thread per particle in a cell, then each thread will calculate the pairwise energy for a particle from three different cells ($9\! \times\! 1\! \times\! 3$). Another example is when only three blocks are called and each thread is calculating one particle from nine different cells ($3\! \times\! 1\! \times\! 9$).

\subsubsection{Cell List Implementation Details}

Incorporating the cell list implementation into the CUDA implementation described in section~\ref{sec:method} requires the modification of two steps. First, in calculating the pairwise energy of the system. For each type of move, only the current cell that a particle belongs to and the 26 neighboring cells'  particles are considered. Second, for each type of move, slightly different steps are required as follows:

Extra steps are required to update the cell list if a particle displacement move is accepted. The list of particles for both the source and the destination cells should be updated, unless the source and destination cells are the same. To remove a particle from the cell list, an exhaustive search for the selected particle index in the source cell is executed. Then the last particle's index in that cell replaces the memory location in which the particle was stored, unless the particle of interest is the last particle in the list. In both cases, the counter of particles for that cell is decremented. The destination cell update requires fewer operations. The particle ID is appended to the destination cell and the counter for that cell is incremented. In addition, the maximum number of particles per cell is used by the CPU so that the correct number of threads per cell are spawned. So, some information is copied from the GPU to the CPU after each move.

\subsection{Microcell List Implementation}\label{subsec:microcellListImp}

Although we show in section~\ref{sec:discussion} that, for large enough problem sizes, a traditional cell list implementation outperforms a GPU implementation without cell lists, there are a few drawbacks to this approach. First, the maximum number of particles in a cell can vary with system parameters such as density and cutoff, so we either use extra memory to store each cell's data or we need to make run-time adjustments based on input parameters, which can make code optimization difficult. Second, efficient load balancing of the threads is difficult, because the number of particles in a cell could differ significantly across cells. Finally, the 27 cells encompass a volume significantly larger than the volume of interest. For example, with a cutoff of $2.5\sigma$ and a cell size of $2.75\sigma$ per dimension, the sphere surrounding a particle has a volume of $65.45$, but the volume of the corresponding cells is more than $561.5$. In general, the cell dimensions will exceed the cell size by a small amount because the simulation box must be subdivided into cells of the same size. Even if an exact cutoff of $2.5\sigma$ is possible, the corresponding volume is $421.875$, which means less than $16\%$ of the volume being processed in the cells could hold a particle within the cutoff.

Dividing the volume of the box into smaller cells, for instance, with a cell size of $r_{cut}/2$, partially addresses some of these drawbacks, but not all of them. Instead, we propose a novel and highly efficient cell list structure that shows \emph{better performance for all problem sizes}. The simulation box is partitioned into microcells, each of which has a volume of $\sigma^3$, except for cells on the boundary, which could be smaller. For instance, if the volume of the simulation box is $50.23\sigma^3$, then the last cell along an axis will be $0.23\sigma$ in that dimension. We then assign each microcell to a unique thread. This organization has several advantages:

\begin{itemize}
\item The total volume being processed is smaller. Even allowing for boundary cases, only $343\ (7 \times 7 \times 7)$ threads are needed for a cutoff of $2.5\sigma$ or $2.75\sigma$.
\item The microcell that contains a particular particle can be computed, without any division, directly from the integer portion of each coordinate.
\item The number of threads per block is fixed and independent of the number of particles in the box, so we do not need to copy results to the CPU prior to each kernel call.
\item The specific microcell assigned to each thread in the cube of cells can be mapped directly from the thread IDs by defining an $n \times n \times n$ dim3D block of threads for the kernel call.
\item Because of the physical properties of the system, we can guarantee that there cannot be more than five particles in a cell. So, we can bound the maximum cell contents regardless of the input parameters.
\item The load balancing is straightforward. For typical densities of 0.5 or 0.6 particles per microcell, we expect on average about half the microcells to have no particles and the other half to have one particle. So, processing is relatively fast.
\item Since only one thread is used per microcell, the entire move runs in one block, eliminating all inter-block synchronization.
\end{itemize}

\subsubsection{Building the Initial Microcell List}

The initial cell list is created when the system, including the particle positions, is initialized. There are two arrays used for the cell list. One contains a counter of the number of particles in a cell. This array is initialized to zero using {\sf cudaMemset()} and updated when a particle is added to a cell. There is also an array that contains the particle numbers, indices into the array of particle coordinates. This array can hold a maximum of five particles per cell. To improve memory coalescing, the array is organized so that we store the first particle of each cell, followed by the second, etc. In other words, if there are $1000$ cells, then array location $0$ holds the index of the first particle in cell $0$, array location $1000$ holds the index of the second particle in cell $0$, etc. Atomic operations are used when adding a particle to a cell. However, note that since the cells are small and the density is less than $1.0$, there is typically at most one particle in a cell at a time, so an atomic operation by one thread does not block another thread, since they are accessing different array locations.

\subsubsection{Assigning Microcells to Threads}

As mentioned above, the particle coordinates are used to compute its cell. The thread ID has three components: threadIdx.x, threadIdx.y, and threadIdx.z. Each corresponds to a unique cell in the cube of microcells, with the specific cell determined as an offset in each dimension from the cell of the particle. For instance, when processing a particle in cell $(7, 6, 5)$ and using a $7 \times 7 \times 7$ cube of microcells, cell $(4, 3, 2)$ is  assigned to thread $(0, 0, 0)$ and cell $(10, 9, 8)$ is assigned to thread $(6, 6, 6)$.  These simulations use periodic boundary conditions, so the code wraps around the box if necessary. The maximum block size used is $512\ (8 \times 8 \times 8)$ threads. For cutoffs greater than $3.75\sigma$, there are more than $512$ microcells, so the threads iterate over multiple microcells. This allows the code to support arbitrarily large cutoffs. There is some loss of efficiency, so a different implementation could be considered for problems that use a cutoff $\ge 4.0\sigma$, although this is uncommon.

\subsubsection{Microcell List Implementation Details}

The CPU determines which move to perform: displacement, insertion, or deletion. The CPU then invokes the appropriate GPU kernel. The GPU uses a single block to evaluate the move and either accept or reject the move. Results do not need to be exchanged between the CPU and GPU except for periodic checkpointing of the system status and at the end of the simulation, except for some general statistics so that properties such as the average energy and average pressure can be calculated.

When a move is accepted, a particle ID may need to be inserted into a microcell and a particle ID may need to be removed from a microcell. Inserting an ID is straightforward; the location where the ID is inserted into the particle array is determined by the number of particles already in the cell and then the number of particles in the cell is incremented. Deleting an ID is simple when there is only one particle in a microcell; just decrement the particle counter for that cell. Because the densities are generally less than $1.0$, this is the typical case. If there are multiple particles in a cell, then we need to scan through the IDs in the list and replace the deleted particle with the ID of the last particle in the list. Of course, this will still be faster than with the traditional cell list, since there are fewer particles in a microcell.

\subsubsection{Further Optimizations}

As shown in section~\ref{sec:discussion}, this novel cell list implementation offers significant performance improvements over both the code without cell lists and the version that uses traditional cell lists. Now, we describe further modifications to the code to fully exploit the advantages of the microcell list properties. To provide a fair comparison, we retained both the optimized and unoptimized versions and report results for both versions. Note that the unoptimized version uses a slightly different design, where only one microcell is assigned to each thread, unlike what is described above. So, the unoptimized version is inherently more efficient for some larger cutoffs, but cannot process cutoffs larger than $4.25\sigma$. Even so, the optimized version is faster than the unoptimized version for all cutoffs. The primary optimizations are:

\begin{itemize}
\item Since the block size is fixed, we create a single kernel call that iterates over multiple moves. This is based on an input parameter that specifies how often checkpointing of the system is done. For the reported results, this corresponds to 10,000 moves. So, we are replacing 10,000 kernel calls with one kernel call. Consequently, things like average energy and average pressure are now computed on the GPU.
\item Since only a single block is used for each move, no global memory is used for storing the results of the energy calculations. Instead, these are returned using parameters.
\item The {\sf \_\_launch\_bounds\_\_()} compiler directive is used to provide a hint to the compiler that we are running just one block of at most $512$ threads. This enables more compiler optimization of the code.
\item The {\sf cudaDeviceSetCacheConfig()} function is used to allocate more L1 cache and less shared memory for all the functions except {\sf CalculateTotalEnergy()}. Since these other functions have just one block, dedicating more on-chip memory to cache improves performance.
\end{itemize}

As shown in the next section, these optimizations, along with many minor tweaks to the code, have resulted in significant performance improvements. Almost all of these optimizations were further improvements that were enabled by the use of the microcell list implementation.


\section{Performance Results}\label{sec:discussion}

In this section, we present a two-part performance evaluation of the cell list implementations of the parallel Monte Carlo simulation for the grand canonical ensemble. The first part of the simulation study uses CUDA toolkit 4.2~\cite{cuda+programming+guide6.0} and evaluates the end-to-end application wall clock time without traditional cell lists against a single core CPU implementation. The serial and CUDA implementations have been executed on a PC with an Intel i5-2500k CPU that has 8 GB of RAM running Linux kernel build 2.6.32 and compiled to a 64-bit executable with the Intel 13.0.0 compiler. Parallel results are collected from running the code on the same machine using a NVIDIA GeForce GTX 480 graphics card. This card has 1.5 GB of global memory, 15 streaming multiprocessors with 32 cores each, and compute capability 2.0. All code has been compiled with the full optimization flag (-O3). The purpose of this comparison is \emph{not}\/ to compare the performance between a CPU and GPU, since we are using only one CPU core. Instead, it is intended to show that the CUDA code without cell lists is a reasonably efficient implementation upon which the cell list code improves.

All these measurements have used one million simulation steps (move attempts) with a cutoff ($r_{cut}$) of $2.5\sigma$. Although one million simulation steps is enough to show the relative performance of the various codes, it is important to note that for larger problems sizes, scientifically valid results require runs of hundreds of millions or billions of steps. Therefore, it would be reasonable to increase the difference in execution times by two or three orders of magnitude to get a better idea of the potential time savings. Note that valid results are produced by both the serial and the CUDA code, since they are statistically equivalent to published results~\cite{potoff:10914}. The Mersenne twister algorithm has been used to generate the pseudo-random numbers used in these simulations~\cite{Matsumoto+Nishimura-Merstwis:98}.

Table~\ref{tbl:exectimesicc} presents the performance of the sequential grand canonical code and the CUDA code for a number of particles ranging from $2^9$ to $2^{18}$ with a corresponding volume of the simulation box ranging from 853.3 to 436905.6, doubling as the number of particles doubles to maintain a fixed initial density of 0.6 particles per $\sigma^3$. For problem sizes of less than 4096 particles, no speedup has been achieved due to low utilization of the GPU. However, when the simulation box has 4096 particles, the CUDA code begins to outperform the serial code, achieving about 2 times speedup as shown. As the system size increases, more mathematical operations are executed and the speedup of the GPU code continues to increase. For the largest problem size, we see a 15.8 fold speedup. The primary reason smaller systems do not show any speedup is because the kernel call overhead for smaller systems exceeds the gain of parallelism. Since MC simulations move only a single particle at a time, there is not enough arithmetic intensity for small systems. But, as the system size grows, the CUDA code shows more speedup, up to around 16 times for the largest problem size. The large number of particles and the associated arithmetic operations expose enough parallelism to hide the overhead of kernel calls for such large systems.

\begin{table}[!t]
\renewcommand{\arraystretch}{1.3}
\caption{Execution time in seconds for serial and CUDA programs.}
\label{tbl:exectimesicc}
\centering
\begin{tabular}{|r|c|c|c|c|c|}
\hline
\textbf{Particles} & \textbf{Serial Code} &\textbf{CUDA} &\textbf{Speedup} \\
\hline
512 & 2.8 & 22.3 & 0.13\\
1024 & 7.9 & 22.5 & 0.35 \\
2048 & 14.2 & 22.8 & 0.62 \\
4096 & 52.8 & 23.4 & 2.25 \\
8192 & 116.3 & 26.7 & 4.36 \\
16384 & 237.7 & 36.7 & 6.48\\
32768 & 502.2 & 56 & 8.96 \\
65536 & 991.8 &91.6  & 10.83 \\
131072 &2061 & 154.6 & 13.33 \\
262144 &4534.8 & 287 & 15.8 \\
\hline
\end{tabular}
\end{table}


\begin{table}
\renewcommand{\arraystretch}{1.3}
\caption{Legend of blocks, cells, and threads per kernel call.}
\label{tbl:legend}
\centering
\setlength{\tabcolsep}{3pt}
\begin{tabular}{|l|c|c|c|}
\hline {Notation} & {Blocks/Kernel}& {Cells/Block}& {Threads/Particle}\\
\hline 27x1 & 27 & 1 & 1\\
\hline 9x3 &  9  & 3 & 1\\
\hline 3x9 & 3 & 9 & 1 \\
\hline 1x27 & 1 & 27 & 1 \\
\hline 9x1x3 & 9 & 1 & 3\\
\hline 3x1x9 & 3 & 1 & 9\\
\hline 1x1x27 & 1 & 1 & 27 \\
\hline
\end{tabular}
\end{table}


In section~\ref{subsec:cellListImp}, we described our implementation of the traditional cell list. Many different mappings of cells to blocks of the cell list algorithm have been evaluated. Table~\ref{tbl:legend} explains the meaning of the corresponding notation.



From the results in table~\ref{tbl:cell_vs_cuda}, we can see that when a kernel uses either 9 or 27 blocks, the best performance is achieved with a speedup of 8.32 and 8.31, respectively. In the case of 27 blocks, only one cell is handled by each block. The slightly extra overhead of synchronization and reduction for additional blocks may make the $9\times 3$ version faster than the $27 \times 1$ version, although the difference is not statistically significant in our results. For this reason, we use the $27 \times 1$ version, which assigns each cell to a different block, in the remainder of our performance comparisons. A more detailed performance evaluation of these algorithms can be found in~\cite{eyadthesis}.



\begin{table}
\renewcommand{\arraystretch}{1.3}
\caption{Speedup of different cell list implementations over no cell list CUDA code.}
\label{tbl:cell_vs_cuda}
\centering
\begin{tabular}{|l|c|c|c|c|c|c|}
\hline
Particles &	27x1&	9x3	&3x9&	1x27	&9x1x3&	3x1x9\\
\hline
1024&	0.77&	0.78&	0.78&	0.67&	0.65&	0.46\\
2048&	0.78&	0.79&	0.78&	0.69&	0.65&	0.47\\
4096&	0.87&	0.88&	0.85&	0.72&	0.68&	0.49\\
8192&	1.00&	1.00&	0.98&	0.82&	0.78&	0.50\\
16384&	1.43&	1.39&	1.35&	1.13&	1.09&	0.68\\
32768	&2.10&	2.10&	2.06&	1.75&	1.66&	1.04\\
65536&	3.36&	3.36&	3.31&	2.83&	2.65&	1.69\\
131072&	5.42&	5.39&	5.31&	4.61&	4.29&	2.78\\
262144&	8.31&	8.32&	8.20&	7.28&	6.88&	4.99\\
\hline
\end{tabular}
\end{table}



The second part of the simulation study compares the performance of the various GPU implementations of the code. This illustrates the improvement offered by our novel microcell list implementation. These simulations were conducted on two different systems. The first system is the same PC as before, but now running CUDA 6.0~\cite{cuda+programming+guide6.0} on an NVIDIA Tesla K40c. The K40c has 12 GB of global memory, 15 streaming multiprocessors with 192 cores each, and compute capability 3.5. This is the highest-end NVIDIA GPU available when these simulations were run. The second system is a six-core AMD Phenom II X6 1045T CPU that has 12 GB of RAM running CUDA 6.0~\cite{cuda+programming+guide6.0} on an NVIDIA GeForce GTX 550 Ti. The GTX 550 Ti has 1.5 GB of global memory, 4 streaming multiprocessors with 48 cores each, and compute capability 2.1. Both PCs are running Windows 7 and the programs were compiled to 64-bit executables using MS Visual Studio 2010 with full optimization. These two systems will be referred to by their respective GPUs to distinguish between them when discussing the results.

\begin{figure}
\centering
\includegraphics[width=1.0\linewidth]{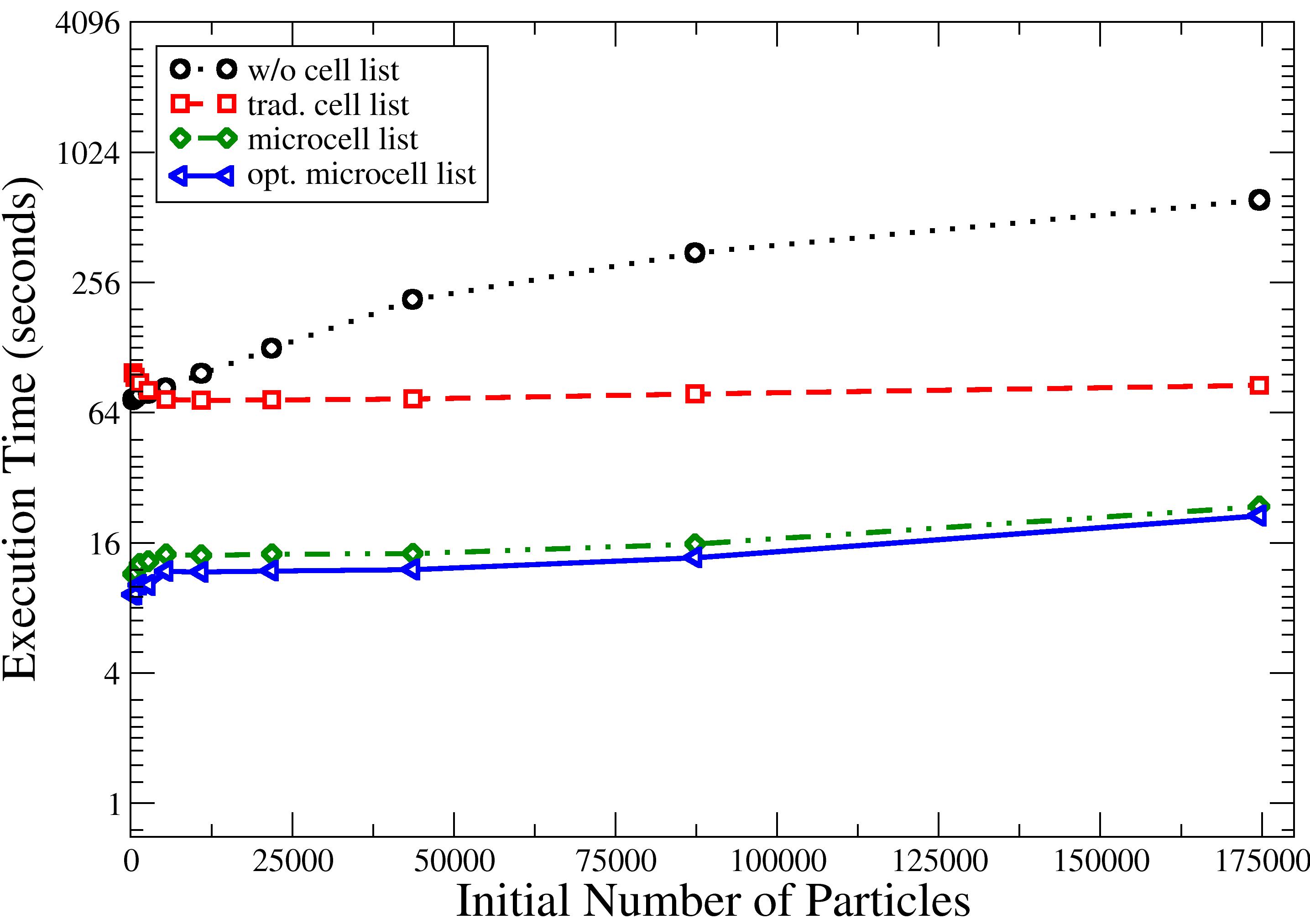}
\caption{Performance of the GPU algorithms on the GTX 550 Ti.}
\label{fig:size550}
\end{figure}

\begin{figure}
\centering
\includegraphics[width=1.0\linewidth]{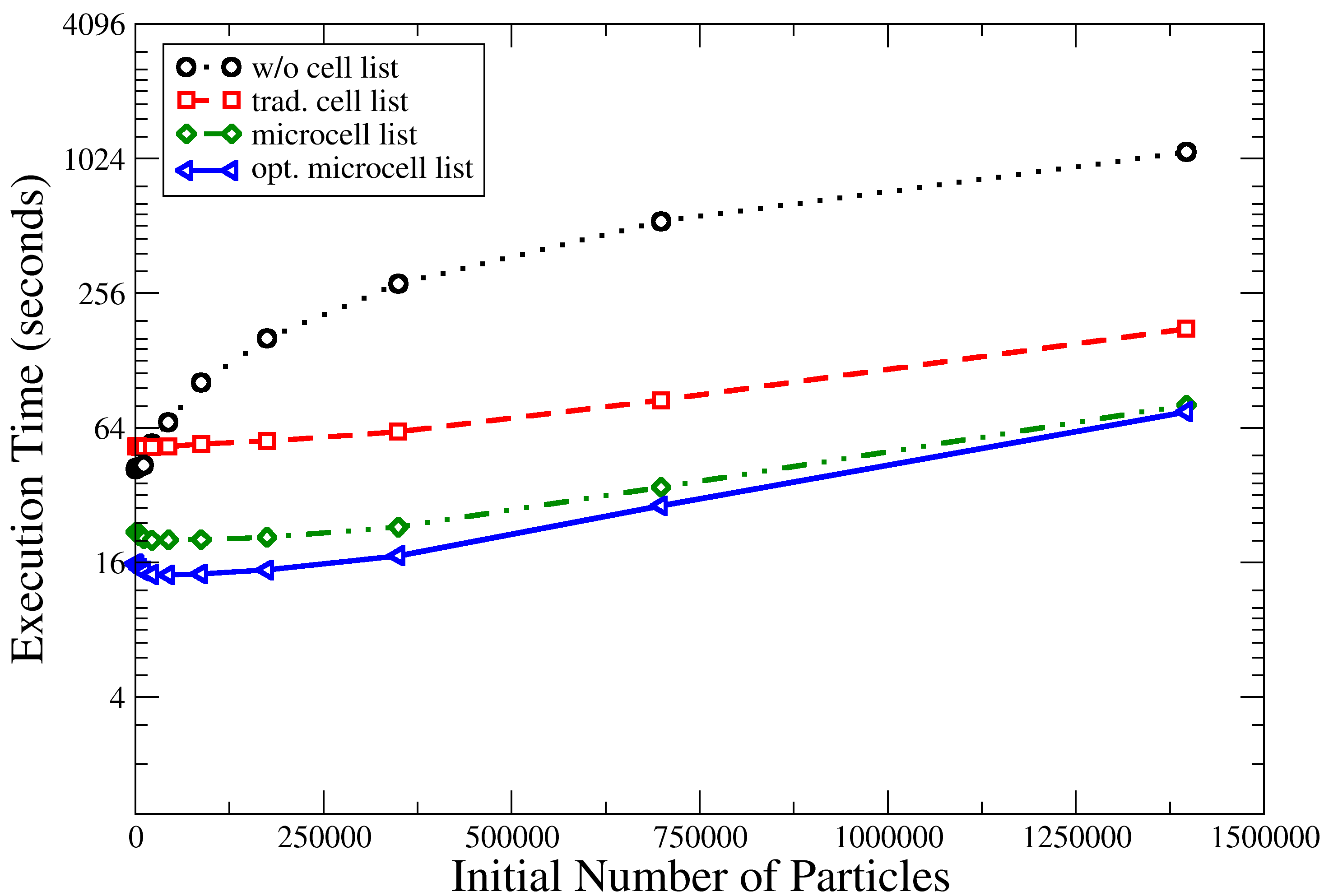}
\caption{Performance of the GPU algorithms on the Kepler K40c.}
\label{fig:sizek40}
\end{figure}

Figures~\ref{fig:size550} and~\ref{fig:sizek40} are log-scale plots that show the results of four different GPU codes: the aforementioned code without cell lists, the aforementioned code with a traditional cell list, a version with the new microcell list, and an optimized version of the microcell list, running on the GTX 550 Ti and the K40c, respectively. All simulations were run for one million steps (move attempts) for different numbers of particles having a fixed initial density of about $0.67$ particles per $\sigma^3$, with a cutoff $(r_{cut})$ of $2.5\sigma$ using a ratio of 30\% displacement moves, 35\% insertions, and 35\% deletions. This initial density was chosen, since longer running experiments (not shown) for these configurations equilibrated at approximately this density. Hence, we able to obtain a better idea of the performance of these codes for long running simulations. Each point in the graphs is the average of five different runs with random seeds. The codes were validated against each other, since the code without cell lists had already been validated against the serial code and the literature. There was little variance in runtime among these five runs, always less than 2\% from the average.

\begin{figure}
\centering
\includegraphics[width=1.0\linewidth]{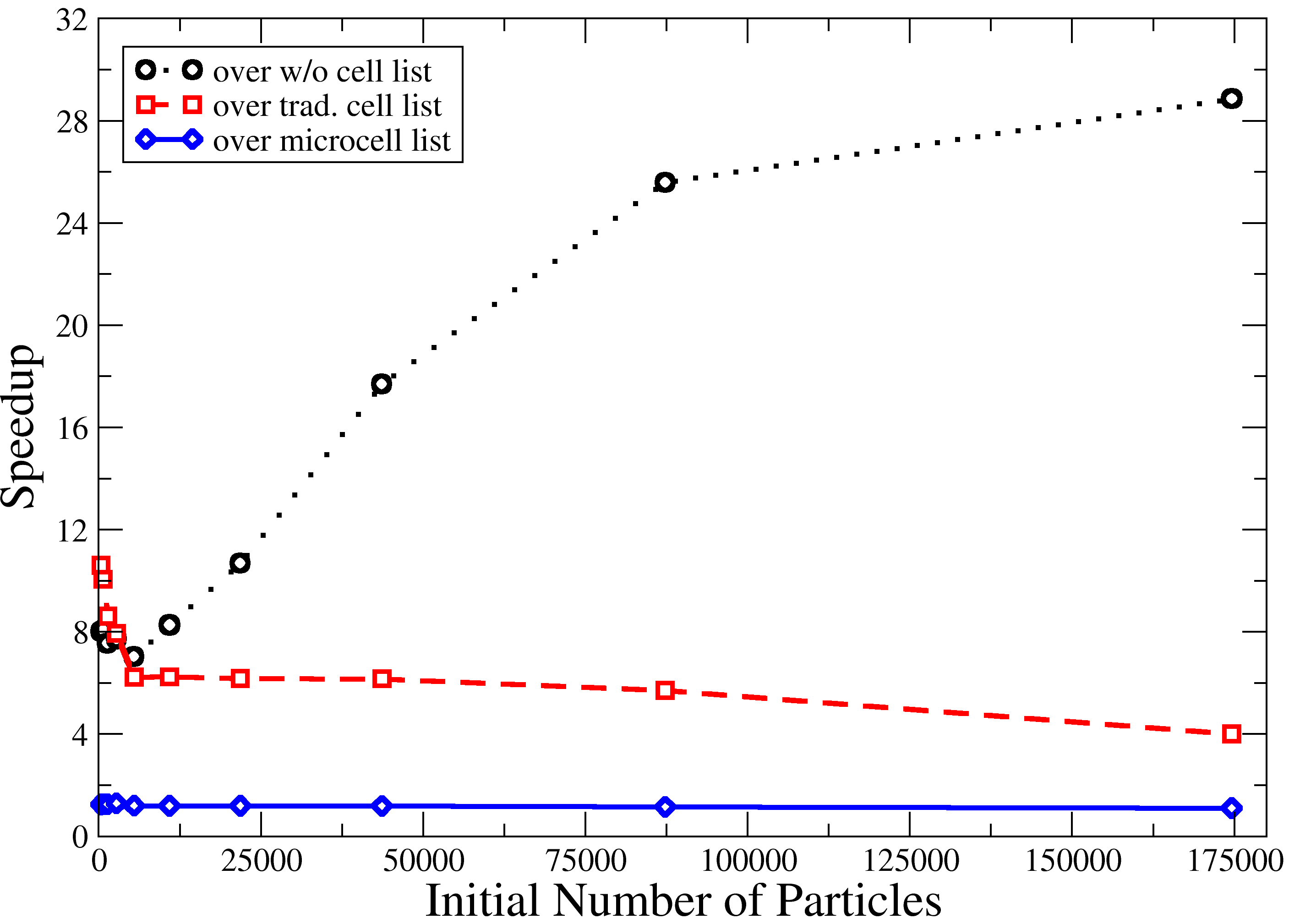}
\caption{Relative speedup of the optimized microcell list on the GTX 550 Ti.}
\label{fig:speedupsize550}
\end{figure}

\begin{figure}
\centering
\includegraphics[width=1.0\linewidth]{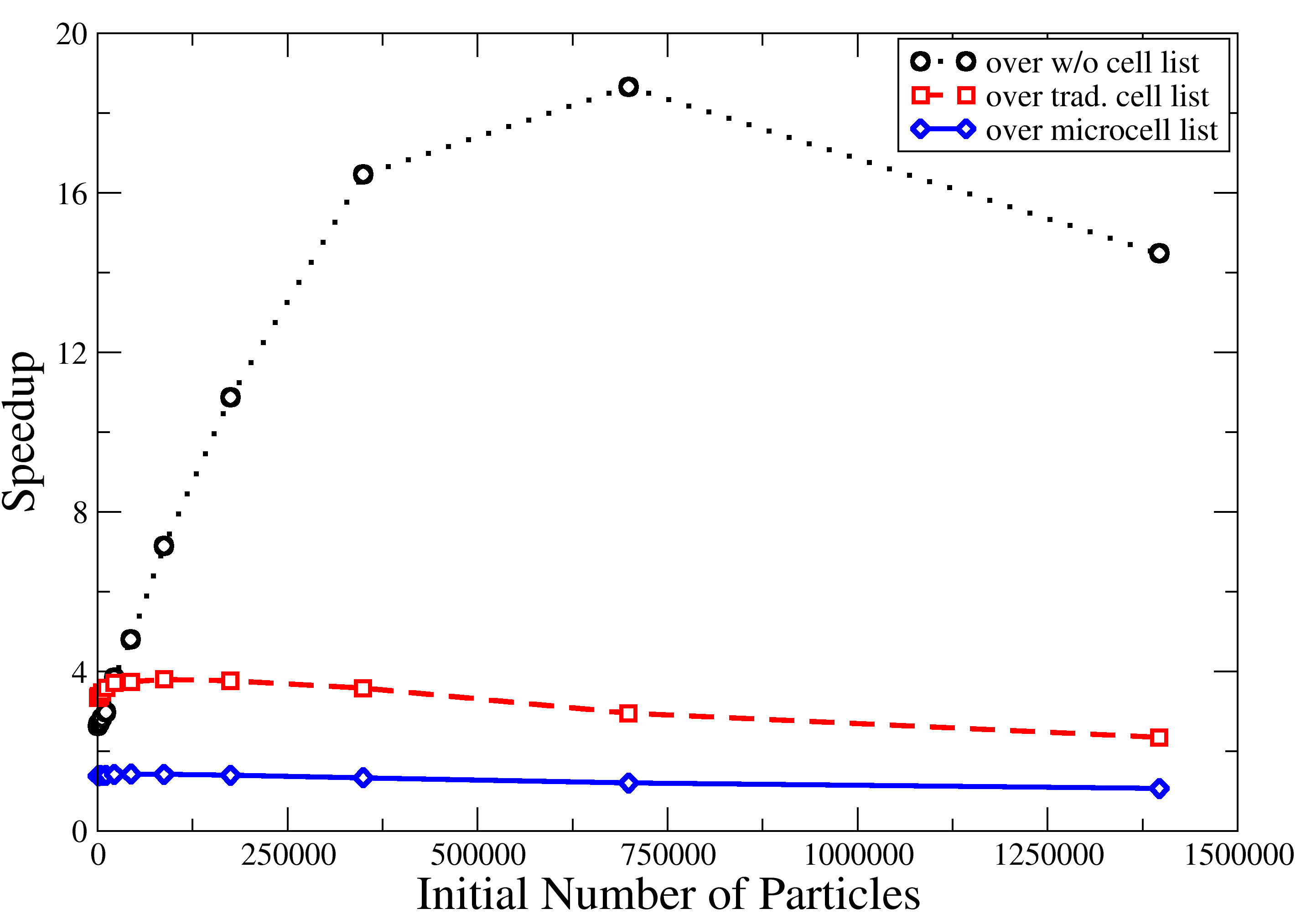}
\caption{Relative speedup of the optimized microcell list on the Kepler K40c.}
\label{fig:speedupsizek40}
\end{figure}

On both systems, the microcell list code shows significantly better performance for all problem sizes. The further optimizations to the microcell list code show additional performance improvement. As shown in figure~\ref{fig:speedupsize550}, the optimized microcell list code running on the GTX 550 Ti is from 8 to 28 times faster than the code without cell lists, with more speedup for larger problem sizes. It is also more than four times faster than the code with a traditional cell list, and between 10\% and 25\% faster than the unoptimized microcell list code. As shown in figure~\ref{fig:speedupsizek40}, on the K40c, which has much more computational resources, the optimized cell list code is up to almost 19 times faster than the code without cell lists, with again more speedup for larger problem sizes. It is also 2 - 4 times faster than the code with the traditional cell list code, and 20\% to 43\% faster than the unoptimized microcell list code, except for the case with over one million particles, when it is only 7\% faster. In the comparisons of the cell list codes, the speedup is more for the smaller problem sizes. The reason is that even though the total system energy is calculated only once, the overhead for this $O(N^2)$ operation becomes an increasingly large portion of the execution time, especially when each move has been optimized to run so quickly.

An interesting observation is that after the system is initialized and the initial system energy is calculated, each subsequent kernel is launched with a single block of 512 threads. Because the GPU does not partition blocks, only a single SM of the GPU is being used. This means that in the case of the K40c, the microcell list code is running all the moves using at most 1/15th the resources of both the version without cell lists and the version with the traditional cell list. Even so, the microcell list code is able to achieve significantly better performance. In essence, the microcell list algorithm allows us to achieve a more effective schedule~\cite{Lutz1995} for computing the energetic decomposition on the GPU. This further demonstrates the potential benefits of using an algorithm tuned specifically to the GPU architecture.

\begin{figure}
\centering
\includegraphics[width=1.0\linewidth]{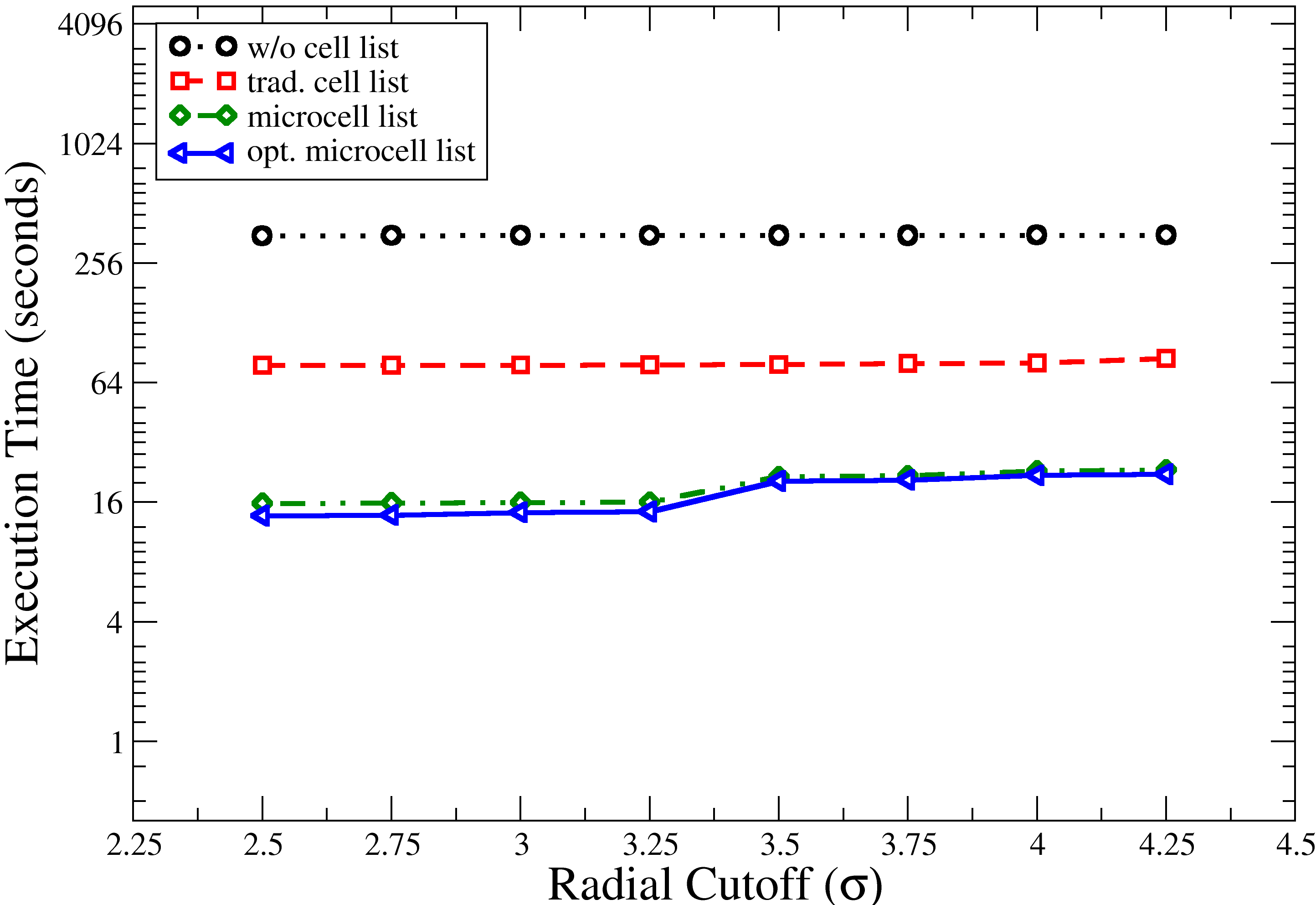}
\caption{Performance impact of different cutoffs with about 85K particles on the GTX 550 Ti.}
\label{fig:cutoff550}
\end{figure}

\begin{figure}
\centering
\includegraphics[width=1.0\linewidth]{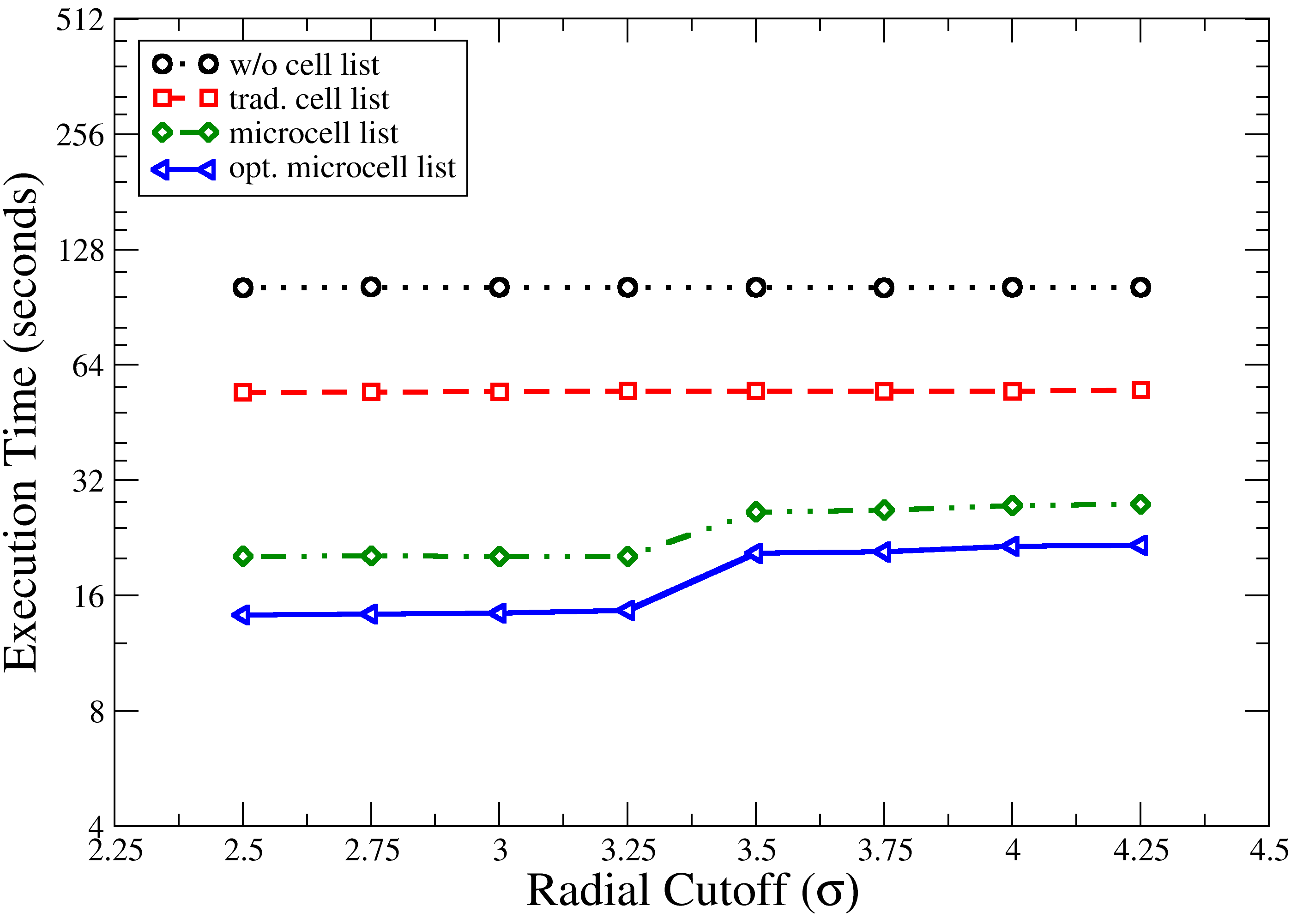}
\caption{Performance impact of different cutoffs with about 85K particles on the Kepler K40c.}
\label{fig:cutoffk40}
\end{figure}

As the cutoff increases, one would expect the relative performance of the cell list codes to decline. To test this hypothesis, we ran simulations with the same parameters as before, except that we fixed the initial configuration to have 87296 particles and a volume of 128K $\sigma^3$, and varied the cutoff from $2.5\sigma$ to $4.25\sigma$. These results are shown as log-scale plots in figures~\ref{fig:cutoff550} and~\ref{fig:cutoffk40}. On both systems, we observe the expected slow decline in the speedup of the optimized microcell list code compared to the other three versions of the GPU code. However, for all cutoffs, the optimized microcell list code remains faster than the other three codes. So, even for systems with larger cutoffs, the microcell list code is a more efficient algorithm.

\section{Conclusions and Future Work}\label{sec:conclusion}

Designing efficient algorithms for a parallel architecture can offer significant performance advantages over just porting an existing sequential algorithm. Our proposed cell list structure is an example of such an algorithm, which provides an efficient mapping of cells to the threads of a manycore architecture. The design of the microcell list allows simple calculations of in which cell a particle particle resides and which neighboring cells need to be processed. Our future work will examine alternative ways of choosing the neighboring cells and eliminating cells beyond the cutoff in order to further reduce the number of microcells that our functions consider.

We first compare an optimized serial code that runs on a single CPU core with our straightforward GPU implementation to show that the GPU implementation without cell list has reasonable performance. We then compare three different implementations of cell list on the GPU: an implementation using traditional cell lists, an implementation using our new microcell list, and an optimized version of the microcell list code.

The microcell list code shows significant performance improvements for all problem sizes even for large cutoffs. The further optimizations that this algorithm allows produce an even more efficient code. Besides achieving superior performance, the microcell list structure simplifies the computations and eliminates enough unnecessary comparisons to allow us to achieve this performance using just one of the streaming multiprocessors (SM) on the GPU. This offers the opportunity to run multiple large-scale simulations on a single GPU simultaneously. Another option is to run multiple moves concurrently, each on a separate SM, and keep the first one that is accepted. We will investigate both of these options in our future work.

The grand canonical ensemble has only three types of moves, and each of those moves changes at most one particle in the system. For this reason, the microcell list evolves slowly during the simulation. For some other ensembles, such as the Gibbs ensemble~\cite{Parallel-Gibbs-ensemble-simulations1995}, volume transfer moves allow the volume of the simulation boxes to increase or decrease, which requires the pairwise energy of all the particles to be recalculated. In addition, if the volume changes, then the cell list needs to be rebuilt because all of the particles in the box move.

Since recalculation of the system energy and pressure are relatively frequent requirements with the Gibbs ensemble, we plan to develop an efficient microcell list version of the {\sf CalculateTotalEnergy()} function. Our future work will also investigate how to rebuild the cell list efficiently. Although this adds overhead to the simulation, it is likely that the microcell list will still offer significant performance improvement. Volume moves are not common and only about half of volume move attempts are accepted. So, the volume changes, on average, about once every 200 moves. This will allow us to amortize the cost of rebuilding the microcell list over many moves.

\section{Acknowledgments}
The authors thank Micah Bojrab for detailed feedback that significantly improved the presentation of the paper. This material is based upon work supported by the National Science Foundation under Grant No.\ CBET-0730768 and OCI-1148168, and Wayne State University's Research Enhancement Program (REP). We also thank NVIDIA for donating most of the graphics cards used in this study.

%
\bibliographystyle{abbrv}
\bibliography{microcell}  

\begin{thebibliography}{10}

\bibitem{hoomd-blue2012}
{HOOMD-blue web page}.
\newblock \url{http://codeblue.umich.edu/hoomd-blue}, Nov 2012.

\bibitem{Allen}
M.~P. Allen and D.~J. Tildesley.
\newblock {\em {Computer Simulation of Liquids}}.
\newblock Oxford University Press, 1987.

\bibitem{Anderson+LorenzETAL-Genepurpmoledyna:08a}
J.~A. Anderson, C.~D. Lorenz, and A.~Travesset.
\newblock General purpose molecular dynamics simulations fully implemented on
  graphics processing units.
\newblock {\em Journal of Computational Physics}, 227(10):5342--5359, May 2008.

\bibitem{TheMetropolisAlgorithm-Beichl:2000}
I.~Beichl and F.~Sullivan.
\newblock {The Metropolis Algorithm}.
\newblock {\em Computing in Science Engineering}, 2(1):65--69, Jan. 2000.

\bibitem{Brugge}
F.~Brug{\`e}.
\newblock {Systolic Calculation of Pair Interactions Using the Cell
  Linked-Lists Method on Multi-processor Systems}.
\newblock {\em Journal of Computational Physics}, 104(1):263--266, Jan. 1993.

\bibitem{Daly20122054}
K.~B. Daly, J.~B. Benziger, P.~G. Debenedetti, and A.~Z. Panagiotopoulos.
\newblock Massively parallel chemical potential calculation on graphics
  processing units.
\newblock {\em Computer Physics Communications}, 183(10):2054--2062, Oct. 2012.

\bibitem{Frenkel+Smit:02}
D.~Frenkel and B.~Smit.
\newblock {\em {Understanding Molecular Simulation: From Algorithms to
  Applications}}.
\newblock Academic Press, San Diego, second edition, 2002.

\bibitem{Grest1989}
G.~S. Grest, B.~D{\"u}nweg, and K.~Kremer.
\newblock {Vectorized link cell Fortran code for molecular dynamics simulations
  for a large number of particles}.
\newblock {\em Computer Physics Communications}, 55(3):269--285, 1989.

\bibitem{eyadthesis}
E.~Hailat.
\newblock {\em {Advanced Optimization Techniques for Monte Carlo Simulation on
  Graphics Processing Units}}.
\newblock PhD thesis, Wayne State University, Aug. 2013.

\bibitem{doi:10.1021/ct200474j}
J.~Kim, J.~M. Rodgers, M.~Ath{\`e}nes, and B.~Smit.
\newblock Molecular monte carlo simulations using graphics processing units: To
  waste recycle or not?
\newblock {\em Journal of Chemical Theory and Computation}, 7(10):3208--3222,
  2011.

\bibitem{Parallel-Gibbs-ensemble-simulations1995}
L.~Loyens, B.~Smit, and K.~Esselink.
\newblock Parallel {G}ibbs-ensemble simulations.
\newblock {\em Molecular Physics}, 86(2):171--183, 1995.

\bibitem{Lutz1995}
D.~Lutz and D.~N. Jayasimha.
\newblock {Do Fixed-Processor Communication-Time Tradeoffs Exist?}
\newblock {\em Parallel Processing Letters}, 5(2):311--320, June 1995.

\bibitem{Matsumoto+Nishimura-Merstwis:98}
M.~Matsumoto and T.~Nishimura.
\newblock {Mersenne Twister: A 623-dimensionally Equidistributed Uniform
  Pseudo-random Number Generator}.
\newblock {\em ACM Trans. Model. Comput. Simul.}, 8(1):3--30, Jan. 1998.

\bibitem{Rice}
W.~Mattson and B.~M. Rice.
\newblock {Near-neighbor calculations using a modified cell-linked list
  method}.
\newblock {\em Computer Physics Communications}, 119(2-3):135--148, June 1999.

\bibitem{metropolis:1087}
N.~Metropolis, A.~W. Rosenbluth, M.~N. Rosenbluth, A.~H. Teller, and E.~Teller.
\newblock {Equation of State Calculations by Fast Computing Machines}.
\newblock {\em The Journal of Chemical Physics}, 21(6):1087--1092, 1953.

\bibitem{Navarro}
C.~A. Navarro and N.~Hitschfeld.
\newblock {Improving the GPU space of computation under triangular domain
  problems}.
\newblock {\em CoRR}, 2013.

\bibitem{cuda+programming+guide6.0}
NVIDIA.
\newblock {\em {CUDA C Programming Guide}}, 6 edition, Feb. 2014.

\bibitem{Parallel-canonical-Monte-Carlo:Keeffe:2009}
C.~J. O'Keeffe and G.~Orkoulas.
\newblock Parallel canonical {M}onte {C}arlo simulations through sequential
  updating of particles.
\newblock {\em The Journal of Chemical Physics}, 130(13):134109, 2009.

\bibitem{potoff:10914}
J.~J. Potoff and A.~Z. Panagiotopoulos.
\newblock Critical point and phase behavior of the pure fluid and a
  {L}ennard-{J}ones mixture.
\newblock {\em The Journal of Chemical Physics}, 109(24):10914--10920, 1998.

\bibitem{Proctor2013}
A.~J. Proctor, C.~A. Stevens, and S.~S. Cho.
\newblock {GPU-Optimized Hybrid Neighbor/Cell List Algorithm for Coarse-Grained
  MD Simulations of Protein and RNA Folding and Assembly}.
\newblock In {\em Proceedings of the International Conference on
  Bioinformatics, Computational Biology and Biomedical Informatics}, BCB'13,
  pages 633--640, New York, NY, USA, 2013. ACM.

\bibitem{rice:1}
O.~K. Rice.
\newblock {On the Statistical Mechanics of Liquids, and the Gas of Hard Elastic
  Spheres}.
\newblock {\em The Journal of Chemical Physics}, 12(1):1--18, 1944.

\bibitem{stone2007accelerating}
J.~Stone, J.~Phillips, P.~Freddolino, D.~Hardy, L.~Trabuco, and K.~Schulten.
\newblock {Accelerating Molecular Modeling Applications with Graphics
  Processors}.
\newblock {\em Journal of Computational Chemistry}, 28(16):2618--2640, 2007.

\bibitem{van2008harvesting}
J.~van Meel, A.~Arnold, D.~Frenkel, S.~{Portegies Zwart}, and R.~Belleman.
\newblock Harvesting graphics power for {MD} simulations.
\newblock {\em Molecular Simulation}, 34(3):259--266, 2008.

\bibitem{Verlet}
L.~Verlet.
\newblock {Computer ``Experiments'' on Classical Fluids. I. Thermodynamical
  Properties of Lennard-Jones Molecules}.
\newblock {\em Phys. Rev.}, 159(1):98--103, 1967.

\bibitem{Wittenbrink2011}
C.~M. Wittenbrink, E.~Kilgariff, and A.~Prabhu.
\newblock {Fermi GF100 GPU Architecture}.
\newblock {\em IEEE Micro}, 31(2):50--59, March-April 2011.

\bibitem{Yao}
Z.~H. Yao, J.-S. Wang, G.-R. Liu, and M.~Cheng.
\newblock Improved neighbor list algorithm in molecular simulations using cell
  decomposition and data sorting method.
\newblock {\em Computer Physics Communications}, 161(1-2):27--35, 2004.

\end{thebibliography}
%
%
%

\end{document}